\documentclass[11pt,onecolumn,amssymb,nofootinbib]{revtex4}
\usepackage{amsmath, amsthm, amscd, amssymb}
\usepackage{bm}
\usepackage{bbm}

\begin{document}

\title{\bf Placing direct limits on the mass of earth-bound dark matter}
\author{Stephen L. Adler}
\email{adler@ias.edu} \affiliation{Institute for Advanced Study,
Einstein Drive, Princeton, NJ 08540, USA.}

\begin{abstract}

We point out that by comparing the total mass (in gravitational units)  of the earth-moon system, as determined by lunar laser ranging, with the sum of the 
lunar mass as independently determined by its gravitational action on satellites 
or asteroids, and the earth mass, as determined 
by the LAGEOS geodetic survey satellite, one can get a direct measure of the 
mass of earth-bound dark matter lying between the radius of the moon's orbit 
and the geodetic satellite orbit.  Current data show that the mass of such earth-bound dark matter must be less than $4 \times 10^{-9}$ of the earth's mass.   
\end{abstract}

\maketitle

Current interest in dark matter has been heightened by the recent report by  
the DAMA/LIBRA collaboration \cite{DAMA} of evidence for galactic halo dark 
matter, based on their observation of an annual modulation 
signal.   Astrophysical arguments suggest that 
the galactic halo dark matter mass density is around $0.3 ({\rm GeV}/c^2) {\rm cm}^{-3}$, but it is still an open question whether in addition to dark matter 
bound to the galaxy, there may be larger dark matter concentrations bound 
to the sun, and bound to the earth.  The possibility of sun-bound dark matter 
was discussed in an article of Fr\`ere, Ling and Vertongen \cite{frere}, who pointed out that 
local dark matter concentrations in the galaxy may have played a role in the 
formation of the solar system.  Their paper, and the papers of Sereno and Jetzer 
\cite{jetzer}, of Iorio \cite{iorio}, and of Khriplovich and 
Pitjeva \cite{khrip}, use arguments based on planetary orbits to place a limit 
on a local excess of sun-bound dark matter of order  $3 \times 10^5$ times the 
galactic halo dark matter mass density. 
It is also possible that there may be 
further local concentrations of earth-bound dark matter, which if large enough 
could be relevant \cite{adler} for understanding the recently reported \cite{and} 
spacecraft flyby anomaly.   Thus, it would be useful to have a direct method for 
measuring, or at least placing limits on, the mass of earth-bound dark matter.

The aim of this note is to show  that it is possible 
to set a direct limit on the total earth-bound dark matter mass lying between 
the radius of the moon's orbit and the radius of low lying satellite 
orbits, such as that of the accurately monitored  \cite{rubin} LAGEOS 
satellite.  For a satellite of negligible mass in a circular orbit around 
an astronomical body of mass $M$, measurement of the orbit radius $R$ and 
the orbital period $T$ gives the product $GM$ (with $G$ the Newton gravitation constant) by use of the formula 
\begin{equation}\label{eq:gmformula}
GM=\frac{4 \pi^2 R^3}{T^2}~~~.
\end{equation}
Thus, from a measurement of the radius and period of the LAGEOS orbit, 
one gets $GM_{\oplus}$, where we have defined the earth mass $M_{\oplus}$ 
here to include the mass of all earth-bound dark matter lying within the radius of 
the LAGEOS orbit.  Similarly, by a measurement of the orbit and period 
of lunar orbiters close to the moon \cite{mich}, one gets $GM_m$, where 
here we have defined the lunar mass $M_m$ to include the mass of all moon-bound dark 
matter lying within the lunar orbiter radius; we shall assume this moon-bound 
dark matter mass to be negligible for purposes of this analysis.  
An alternative way of independently determining the moon's mass 
is to study the orbit of a near-passing asteroid, such as Eros \cite{eros},  which is influenced by the 
gravitational field of the moon as well as the earth.  From such an analysis one 
can extract an accurate figure for the ratio $R_{\oplus/m}\equiv (GM_{\oplus}+ G\Delta M_{\oplus})/(GM_m +  G\Delta  M_m)$, with $\Delta M_{\oplus}$ and $\Delta M_m$ denoting respectively possible small contributions from earth-bound and moon-bound dark matter.   Again assuming that moon-bound dark matter can be neglected, and 
expanding in the small quantity $G \Delta M_{\oplus}$, this ratio becomes 
\begin{equation}\label{eq:ratio}
R_{\oplus/m}=\frac {GM_{\oplus}} {GM_m} (1+\delta)~~~,
\end{equation}
with the small positive correction $\delta$ given by 
\begin{equation}\label{eq:deltadef}
\delta = \frac {\Delta M_{\oplus}}{M_{\oplus}}~~~. 
\end{equation}

Finally, let us consider the 
orbital system comprising the earth and the moon, for which the combined dynamics of the earth-moon system has to be taken into account.\footnote{The influence of dark matter on the earth--moon system has been 
investigated earlier with the aim of determining whether the gravitational interaction of galactic dark matter with the ordinary matter in the earth and moon obeys the equivalence principle, and the equivalence principle in this context has been verified to high accuracy;  see \cite{nord} for details.   The assumption 
that dark matter has normal gravitational interactions with ordinary matter is implicit in our analysis.}  This shows 
\cite{muller} that measuring the relative earth moon distance by lunar laser 
ranging, together with the moon's orbital period, gives a determination of the combined 
mass in gravitational units of the earth-moon system, which is $GM_{\rm combined} =GM_{\oplus}+GM_m+GM_{\rm dm}$, with $M_{\rm dm}$ now the mass of earth-bound dark matter\footnote{If the earth-bound dark matter distribution is not spherically 
symmetric, then there will be dark-matter contributions to higher multipoles of 
the earth's gravitational potential, as well as to the monopole mass.}   
lying between the radius of the moon's orbit and the radius of the LAGEOS satellite 
orbit.  Thus, by subtracting from $GM_{\rm combined}$, as determined by the 
multi-parameter fit 
to the lunar laser ranging experiment, the values of $GM_{\oplus}$ determined by 
LAGEOS and $GM_{m}$ determined by the lunar orbiters, one can get a direct determination of $GM_{\rm dm}$, 
\begin{equation}\label{eq:sub0}
GM_{\rm dm}= GM_{\rm combined}-GM_{\oplus}-GM_m~~~,
\end{equation} 
subject to our assumption that the moon-bound dark matter lying within the lunar orbiter radius can be neglected.  Alternatively, 
if one uses an asteroid determination of $GM_m$, the subtraction to be performed is 
\begin{equation}\label{eq:sub}
GM_{\rm combined}-GM_{\oplus}-\frac {GM_{\oplus}}{R_{\oplus/m}}
\simeq GM_{\rm dm} + GM_m \delta = GM_{\rm dm} + \frac {M_m}{M_{\oplus}}G\Delta M_{\oplus}
\end{equation}
Since $M_{m}/M_{\oplus} \simeq 0.0123$, \eqref{eq:sub} gives  
\begin{equation}\label{eq:sub1}
GM_{\rm combined}-GM_{\oplus}-\frac {GM_{\oplus}}{R_{\oplus/m}}
\simeq GM_{\rm dm}+0.0123 G \Delta M_{\oplus}>GM_{\rm dm}~~~.
\end{equation}
Moreover, if one assumes  the earth-bound dark matter $\Delta M_{\oplus}$
relevant for the asteroid orbit to be similar in magnitude to the 
earth-bound dark  $M_{\rm dm}$ matter lying between the moon's orbit and the LAGEOS orbit, then \eqref{eq:sub1} becomes  
\begin{equation}\label{eq:sub2} 
GM_{\rm combined}-GM_{\oplus}-\frac {GM_{\oplus}}{R_{\oplus/m}}
\simeq GM_{\rm dm}[1+O(.01)]~~~,
\end{equation} 
giving a determination of $M_{\rm dm}$ with a potential  one percent accuracy,  if 
the quantities on the left hand side of \eqref{eq:sub2} were known to sufficient  
accuracy.  
In fact, with current data, the errors in the left-hand side are more significant 
than the error arising from the unknown term of order 0.01 on the right of \eqref{eq:sub2}.   

Proceeding now to a numerical evaluation,\footnote{The cited papers do not give details about how errors are calculated, but the 
context suggests that they are estimated errors based on residuals to multi-parameter model fits.}   the best evaluation  of $GM_{\oplus}$ from 
LAGEOS data is \cite{Ries} $GM_{\oplus}=398600.4415 \pm 0.0008 {\rm km}^3{\rm s}^{-2}$, which when converted to a TDB compatible (Barycentric Dynamical Time 
compatible) figure is  \cite{slava} $GM_{\oplus}=398600.4356 \pm 0.0008 {\rm km}^3{\rm s}^{-2}$.  For $GM_{\rm combined}$, the lunar ranging fit EP0 in table 1 
of \cite{slava1}, in which the sun/(earth +moon) mass ratio was treated as a solution parameter,  gives $M_{\odot}/M_{\rm combined}=328900.5596\pm0.0011$, 
which  converts \cite{slava} (using $GM_{\odot}=1.32712440018(8) \times 10^{11} {\rm km}^3 {\rm s}^{-2}$)  to $GM_{\rm combined}=403503.2357\pm0.0014 
{\rm km}^3 {\rm s}^{-2}$.  The lunar orbiter measurements reported in \cite{mich} 
give $GM_m=4902.84 {\rm km}^3 {\rm s}^{-2}$, with an uncertainty of around 
$\pm.05 {\rm km}^3 {\rm s}^{-2}$ based on a comparison with alternative determinations, but a much more accurate value is obtained from the Eros ranging data of \cite{eros}, which gives $R_{\oplus/m} = 81.300570\pm0.000005$, which corresponds, using the LAGEOS value for $GM_{\oplus}$, to $GM_m=4902.8000\pm0.0003{\rm km}^3 {\rm s}^{-2}$.  Substituting these numbers into \eqref{eq:sub0} or \eqref{eq:sub2} 
gives 
\begin{align}\label{eq:final}
GM_{\rm dm}\simeq& (403503.2357\pm0.0014-398600.4356\pm0.0008 -4902.8000\pm0.0003){\rm km}^3 {\rm s}^{-2} \cr
=&(0.0001 \pm 0.0016) {\rm km}^3 {\rm s}^{-2}
=(0.3\pm 4)\times 10^{-9} GM_{\oplus}~~~,\cr
\end{align}
with the dominant contribution to the error coming from the error in $M_{\rm combined}$ from the 
lunar laser ranging fit.  
Thus, current data show that the mass of  earth-bound dark matter lying between the 
moon's orbit radius $\sim 384,000$ km and the LAGEOS orbit radius\footnote{The LAGEOS orbit is usually described in terms of its altitude of 
$5,900$ km above the earth's surface, which lies about $6,400$ km from the earth's 
center.} $\sim 12,300$ km must be less than $4 \times 10^{-9}$ of the earth's mass, at a 1 $\sigma$ confidence level.\footnote{If dark matter 
gravitationally bound to the earth were assumed uniformly distributed between the moon's orbit and the LAGEOS orbit, this bound, if saturated, would correspond to a dark matter density of order $10^{10}  ({\rm GeV}/c^2) {\rm cm}^{-3}$, much higher than the galactic halo density or current limits on the density of dark matter gravitationally bound to the solar system. Just based on this, however, 
one cannot make any statements on what should have been seen in dark matter detection experiments, since that would require 
making assumptions about the dark matter density profile around the earth, its mass, and its interaction cross section with 
ordinary matter, all of which enter into determining the experimental sensitivity.    No such model-dependent assumptions enter the purely gravitational analysis given above.}  As the accuracy of lunar laser ranging improves, one can expect this limit on  $M_{\rm dm}$ to improve.

I wish to thank Slava Turyshev for inviting me to speak at the workshop From Quantum to Cosmos -- III, Airlie, VA that he organized, and for a subsequent email giving me the numbers and references 
used in the  numerical evaluation of the preceding paragraph.  I also wish to thank  Peter Bender for a helpful conversation at the Airlie workshop, which was the impetus for this investigation, and J. M. Fr\`ere for email correspondence.     This work was supported in part by the Department of Energy under grant no. 
DE-FG02-90ER40542, and I also wish to acknowledge the hospitality of the Aspen 
Center for Physics.  

{\bf Added Note} Gary Gibbons \cite{gibbemail} has pointed out that if one assumes that there is {\it no} 
dark matter bound to the earth, then the comparison of $GM_{\oplus}$ as determined by LAGEOS, with that determined by lunar ranging, gives a bound on possible non-Newtonian modifications to the gravitational force, and he has alerted me to  several references 
\cite{gibbons}, \cite{rapp}, \cite{hubler} where the use of satellite orbits to 
restrict non-Newtonian force models has been discussed.  To illustrate with the 
numbers employed above in the dark matter discussion, if one assumes  
$G=G_{\rm far}$ for the $G$ value relevant both for lunar ranging and for the asteroid determination of the earth to moon mass ratio, and $G=G_{\rm near}$ for the $G$ value relevant 
for the LAGEOS orbit, and takes $M_{\rm dm}=0$, then one has $ G_{\rm near} M_{\oplus}=398600.4356\pm0.0008{\rm km}^3 {\rm s}^{-2}$, 
$G_{\rm far}(M_{\oplus}+M_{m})=403503.2357\pm0.0014{\rm km}^3 {\rm s}^{-2}$, and $R_{\oplus/m}=M_{\oplus}/M_m=81.300570\pm0.000005$.    When combined these give
\begin{equation}\label{eq:gchange}
(G_{\rm near}-G_{\rm far})/G=(0.2\pm 4)\times 10^{-9}~~~,
\end{equation}
indicating that $G$ can change by at most $\sim 4 \times 10^{-9}$ (the same fractional error 
that appears in \eqref{eq:final}) between the radius of the LAGEOS orbit and the radius of the moon's orbit.   This is a factor of five 
better than the result given  
given some time ago by Rapp \cite{rapp}. However,  Turyshev \cite{turyshev}, in reviewing fits to 
lunar ranging, which model the earth-moon distance to 4 mm accuracy, notes that ``analysis of the LLR data tests the gravitational 
inverse-square law to $3\times 10^{-11}$ of the gravitational field strength on scales of the Earth-moon distance''.


\begin{thebibliography}    {17}

\bibitem{DAMA}
R. Bernabei, P. Belli, F. Cappella, R. Cerulli, C. J. Dai, A.
d'Angelo, H. L. He, A. Incicchitti, H. H. Kuang, J. M. Ma, F.
Montecchia, F. Nozzoli, D. Prosperi, X. D. Sheng, and  Z. P. Ye,
``First results from DAMA/LIBRA and the combined results with
DAMA/NaI'', arXiv:astro-ph/0804.2741.


\bibitem{frere} J.-M. Fr\`ere, F.-S. Ling, and G. Vertongen,
 Phys. Rev. D {\bf 77}, 083005 (2008).
 
\bibitem{jetzer} M. Sereno and Ph. Jetzer, Mon. Not. R. Astron. Soc. 
{\bf 371}, 626 (2006).  

\bibitem{iorio} L. Iorio, JCAP 0605, 002  (2006).   
 
\bibitem{khrip}
I. B. Khriplovich and E. V. Pitjeva,  Int. J. Mod. Phys. D
{\bf 15}, 615 (2006); I. B.  Khriplovich,  Int. J. Mod. Phys. D
{\bf 16}, 1475 (2007).

\bibitem{adler} S. L. Adler,  ``Can the flyby anomaly be attributed to earth-bound 
dark matter?'', arXiv:0805.2895. 


\bibitem{and}
J. D. Anderson, J. K. Campell, J. E. Ekelund, J. Ellis, and J. F.
Jordan, Phys. Rev. Lett. {\bf 100}, 091102 (2008).


\bibitem{rubin} D. P. Rubincam, J. Geophys. Res. {\bf 95}, 4881 (1990); 
R. Scharroo, K. F. Wakker, B. A. C. Ambrosius, and R. Noomen, J. Geophys. Res.  
{\bf 96}, 729 (1991).

\bibitem{mich} W. H. Michael and W. T. Blackshear, ``Recent Results on the 
Mass, Gravitational Field and Moments of Inertia of the Moon'', in {\it The Moon 3},  D. Reidel, Dordrecht, pp. 388-402 (1972); available from the NASA Astrophysics 
Data System.    

\bibitem{eros} A. S. Konopliv, J. K. Miller, W. M. Owen, D. K. Yeomans, and J. D. Giorgini, Icarus {\bf 160}, 289 (2002).  

\bibitem{nord} K. L. Nordtvedt, Astrophys. Journ. {\bf 437}, 529 (1994); 
K. L. Nordtvedt, J. M\"uller, and M. Soffel, Astron. Astrophys. {\bf 293}, L73 (1995). 

\bibitem{muller} J. M\"uller, J. G. Williams, and S. G. Turyshev,
``Lunar Laser Ranging Contributions to Relativity and Geodesy'',
arXiv:gr-qc/0509114.

\bibitem{Ries} J. C. Ries, R. J. Eanes, C. K. Shum, and M. M. Watkins, 
Geophys. Res. Lett. {\bf 19}, 529 (1992).  

\bibitem{slava}  S. G.  Turyshev, Jet Propulsion Laboratory, private email communication.  The quoted value 
of $GM_{\odot}$ comes from the NASA/JPL website: http://ssd.jpl.nasa.gov/?constants   

\bibitem{slava1}  J. G. Williams, S. G. Turyshev, and D. H. Boggs, 
``Lunar Laser Ranging Tests of the Equivalence Principle with the Earth and 
Moon'', arXiv:gr-qc/0507083.    

\bibitem{gibbemail}  G. W. Gibbons, private email communication. 

\bibitem{gibbons}  G. W. Gibbons and B. F. Whiting, Nature Lett. {\bf 291}, 636 (1981).  

\bibitem{rapp} R. H. Rapp, Geophys. Res. Lett. {\bf 14}, 730 (1987). 

\bibitem{hubler} B. Hubler, A. Cornaz, and W. K\"undig, Phys. Rev. D {\bf 51}, 4005 
1995).  

\bibitem{turyshev} S. G. Turyshev, ``Experimental Tests of General Relativity'', 
arXiv:0806.1731.   

\end{thebibliography}
\end{document}